\documentclass[a4paper,11pt]{article}
\pdfoutput=1 

\usepackage{jinstpub} 

\newcommand{\gx}{\textsc{GlueX}}

\title{\boldmath The \gx{} DIRC program}

\author[b]{A. Ali,}
\author[d]{F. Barbosa,}
\author[e]{J. Bessuille,}
\author[d]{E. Chudakov,}
\author[b]{R. Dzhygadlo,}
\author[d,e]{C. Fanelli,}
\author[c]{J. Frye,}
\author[e]{J. Hardin,}
\author[f]{A. Hurley,}
\author[a]{G. Kalicy,}
\author[e]{J. Kelsey,}
\author[f]{W. Li,}
\author[e]{M. Patsyuk,}
\author[b]{C. Schwarz,}
\author[b]{J. Schwiening,}
\author[c]{M. Shepherd,}
\author[f,1]{J. R. Stevens,\note{Corresponding author.}}
\author[d]{T. Whitlatch,}
\author[e]{M. Williams,}
\author[e]{Y. Yang,}

\affiliation[a]{Catholic University of America\\Washington DC, United States}
\affiliation[b]{GSI Helmholtzzentrum für Schwerionenforschung GmbH,\\Darmstadt, Germany}
\affiliation[c]{Indiana University,\\Bloomington, IN, United States}
\affiliation[d]{Jefferson Lab,\\Newport News, VA, United States}
\affiliation[e]{Massachusetts Institute of Technology,\\Cambridge, MA, United States}
\affiliation[f]{William \& Mary\\Williamsburg, VA, United States}

\emailAdd{jrstevens01@wm.edu}

\abstract{The \gx{} experiment is located in experimental Hall D at Jefferson Lab (JLab) and provides a unique capability to search for hybrid mesons in high-energy photoproduction, utilizing a $\sim$9 GeV linearly polarized photon beam. The initial, low-intensity phase of \gx{} was recently completed and a high-intensity phase has begun in 2020 which includes an upgraded kaon identification system, known as the DIRC (Detection of Internally Reflected Cherenkov light), utilizing components from the decommissioned BaBar DIRC. The identification of kaon final states will significantly enhance the \gx{} physics program, to aid in inferring the quark flavor content of conventional (and potentially hybrid) mesons.  In these proceedings, we describe the installation of the \gx{} DIRC and the analysis of initial commissioning data.}

\proceeding{International Workshop on Fast Cherenkov Detectors - Photon detection, DIRC design and DAQ\\
11 - 13 September 2019,\\ 
Castle Rauischholzhausen, Justus-Liebig-University Giessen, Germany}

\begin{document}
\maketitle
\flushbottom

\section{Introduction}
\label{sec:intro}

\begin{figure} [h]
    \begin{center}
        \includegraphics[width=0.8\textwidth]{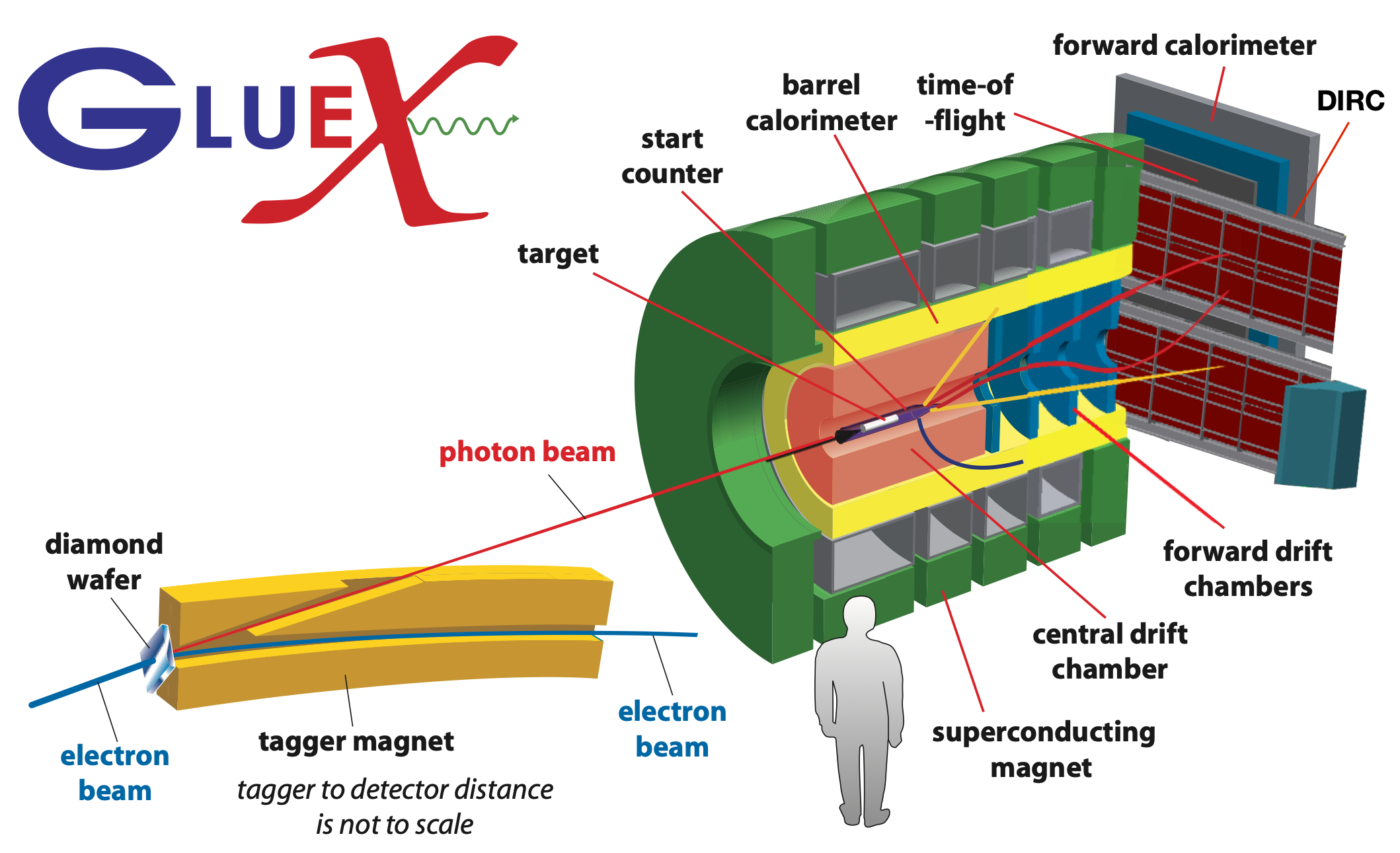}
    \caption{A schematic of the Hall D beamline and \gx{} detector at Jefferson Laboratory.  The DIRC detector is installed directly upstream of the time-of-flight detector in the forward region.}
\label{fig:GlueXcartoon}
\end{center}
\end{figure}

The \gx{} experiment, shown schematically in Fig.~\ref{fig:GlueXcartoon} and located in Jefferson Lab's Hall D, utilizes a tagged photon beam derived from the electron beam's coherent bremsstrahlung radiation from a thin diamond wafer.  The primary goal of the experiment is to search for and ultimately study an unconventional class of mesons, known as hybrids mesons, which contain an intrinsic gluonic component to their wave functions~\cite{PAC30,PAC40,PAC42}.  Hybrid meson states are predicted by Lattice QCD calculations~\cite{Dudek:2013yja}, and provide an opportunity to quantitatively test our understanding of the strong nuclear force in this non-perturbative regime.

Construction and installation of the baseline \gx{} detector were completed in 2014, with first physics-quality photon beams delivered in Spring 2016~\cite{Ghoul:2015ifw}.  The low-intensity phase of the physics program with the baseline \gx{} detector then ran from 2017-2018.  The particle identification (PID) capabilities have been studied with this data, and in the forward region the time-of-flight (TOF) detector performance has reached its design specifications for providing $\pi/K$ separation up to $p\sim2$~GeV$/c$.  An initial physics program to search for and study hybrid mesons which decay to non-strange final state particles is underway.  An upgrade to the PID capabilities is needed to fully exploit the discovery potential of the \gx{} experiment by studying the quark flavor content of the potential hybrid states.  The DIRC upgrade for \gx{}, described in detail in Ref.~\cite{dirc_tdr}, utilizes one-third of the fused silica radiators from the BaBar DIRC (Detection of Internally Reflected Cherenkov light) detector~\cite{Adam:2004fq}, with two new, compact expansion volumes.  These proceedings describe the installation of the \gx{} DIRC and the analysis of initial commissioning data.

\section{Installation}
\label{sec:install}

\begin{figure} [h]
    \begin{center}
        \includegraphics[width=0.48\textwidth]{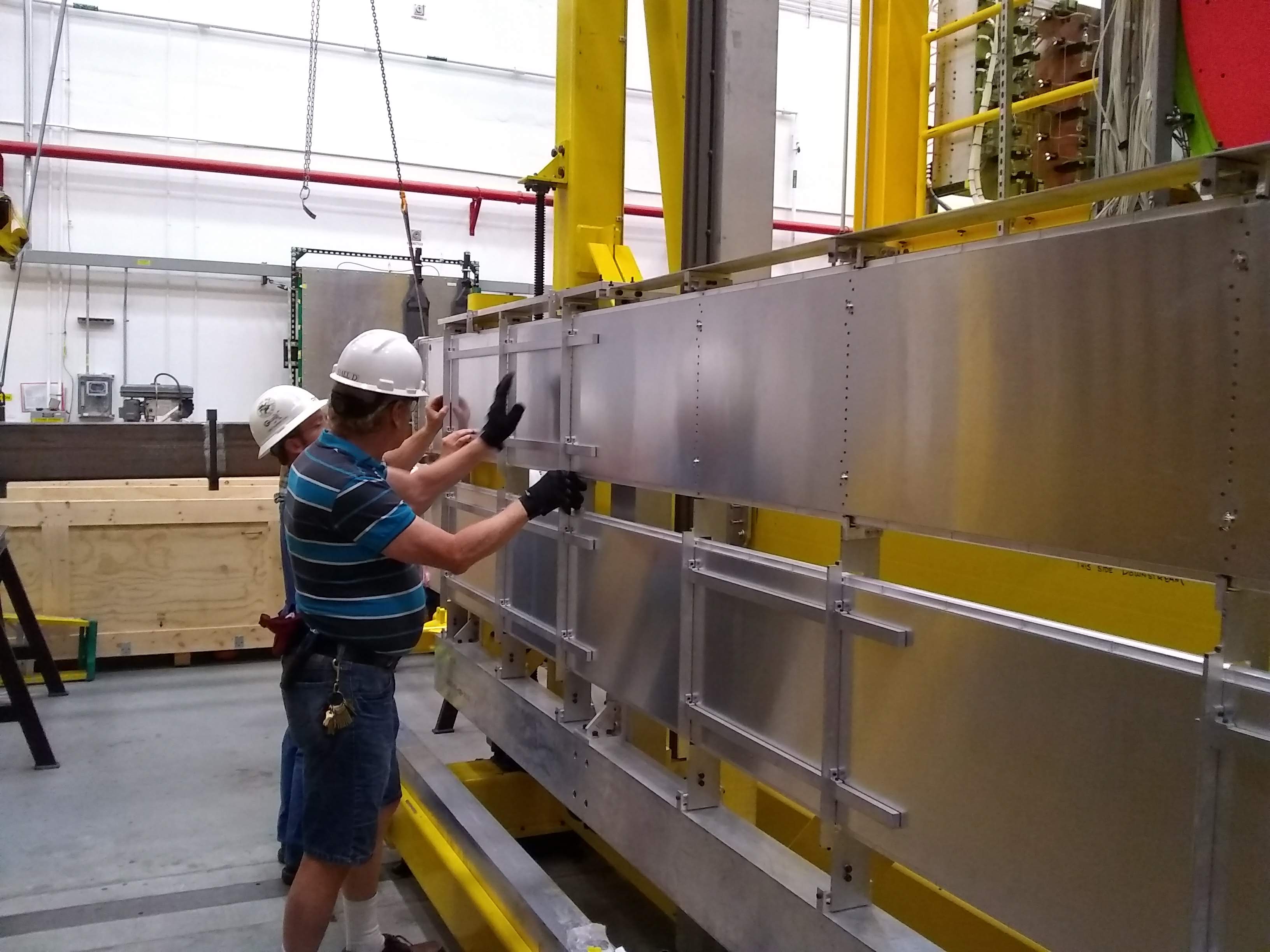}
        \includegraphics[width=0.48\textwidth]{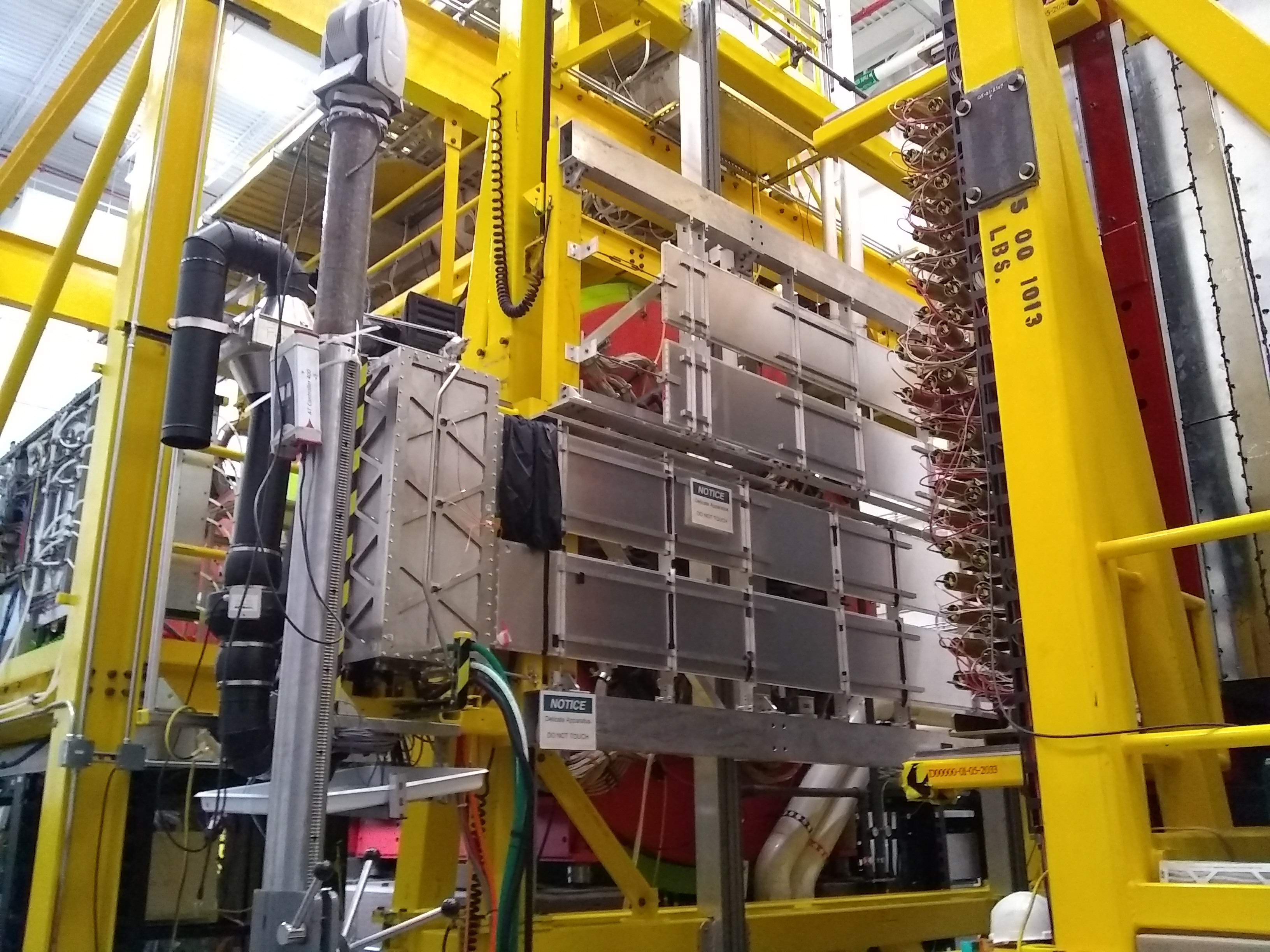}
    \caption{Four BaBar bar boxes transported to JLab and installed in the \gx{} detector on the DIRC support structure in Hall D.}
\label{fig:barbox}
\end{center}
\end{figure}

The first major milestone of the ~\gx{} DIRC installation was successfully transporting four of the DIRC ``bar boxes" from SLAC to JLab in 2018, which each are recycled components from the BaBar detector consisting of $\sim5$~m long fused silica radiator bars, which are quite fragile, contained in an aluminum housing~\cite{Patsyuk:2020wnh}.  Once at JLab, these bar boxes were installed in their support structure for operation in Hall~D, as shown in Fig.~\ref{fig:barbox}.

\begin{figure} [h]
\begin{center}
    \includegraphics[width=0.3\textwidth]{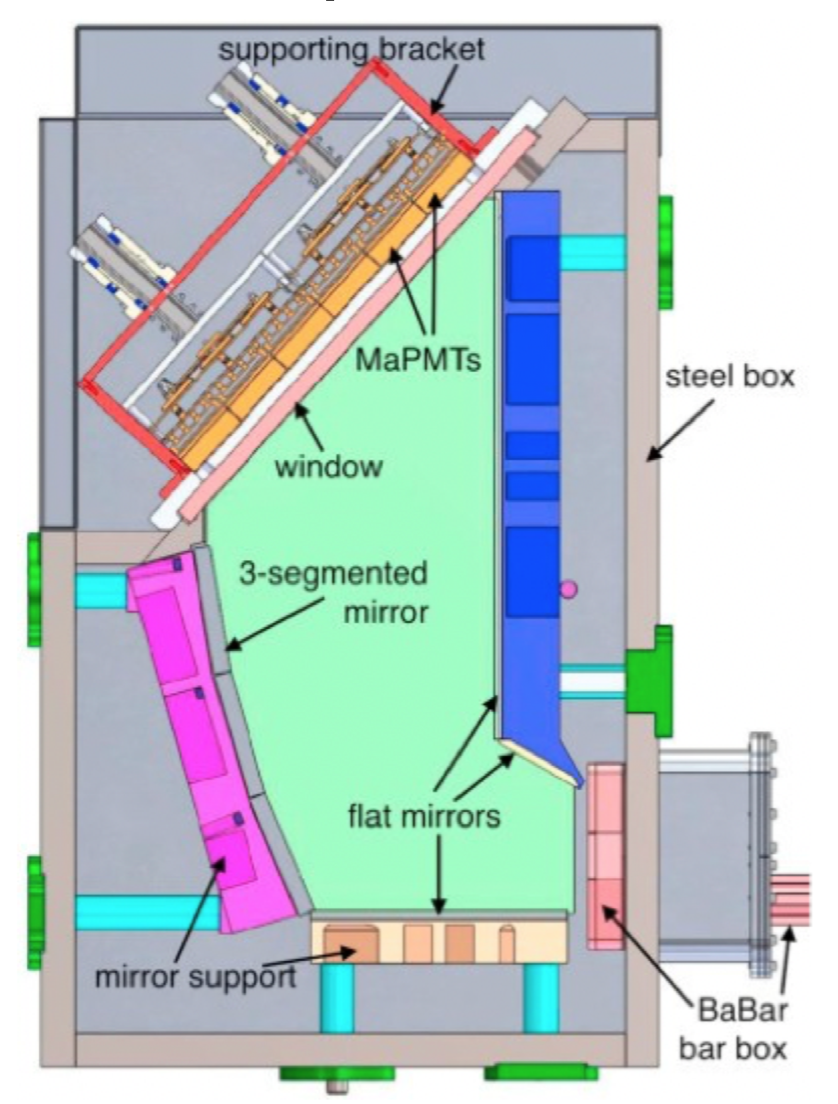}
    \includegraphics[width=0.249\textwidth]{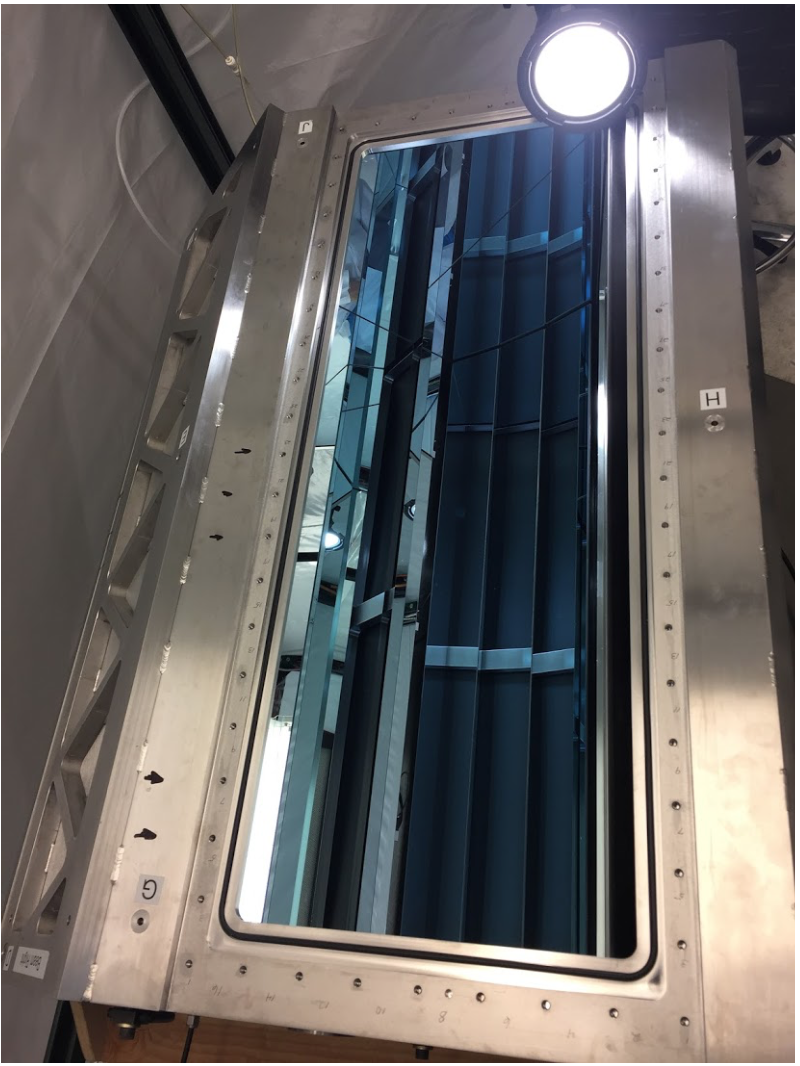}
    \includegraphics[width=0.44\textwidth]{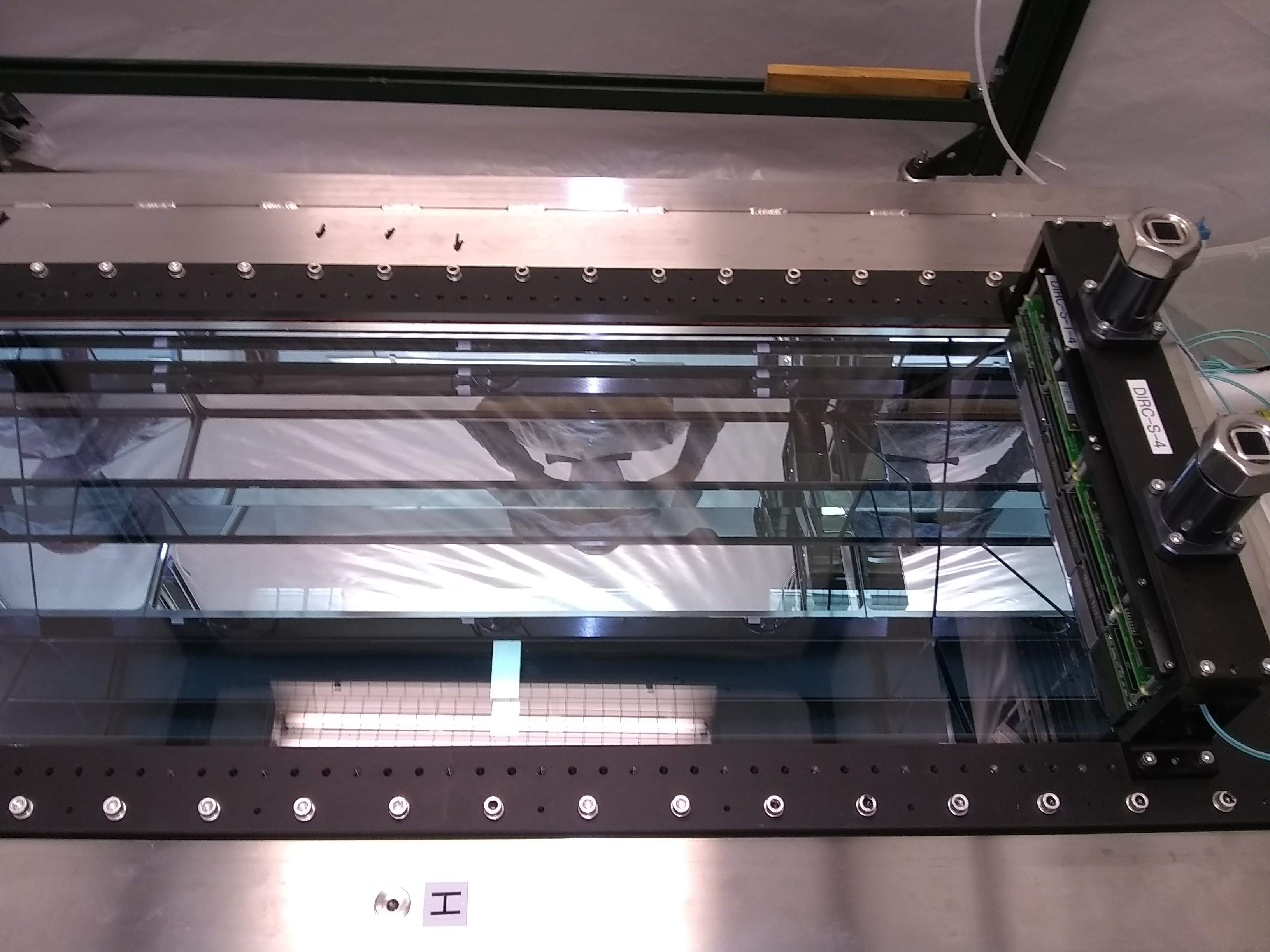}
    \caption{Optical Box design and components (left), mirror components installed in optical box (center), and fused silica window with installed MAPMT modules (right).}
    \label{fig:opticalbox}
\end{center}
\end{figure}

The remaining optical components for the DIRC, known as the Optical Box (OB), consists of multiple flat mirrors, immersed in water, that reflect the Cherenkov photons through a fused silica window to a plane of Multi-Anode PMTs (MAPMTs) as shown in Fig.~\ref{fig:opticalbox}.  The front surface flat mirrors\footnote[1]{https://firstsurfacemirror.com} were characterized for their reflectivity vs wavelength to be utilized in simulation.  The characterization and calibration of the 64-channel Hamamatsu H12700 MAPMTs~\footnote[2]{https://www.hamamatsu.com/jp/en/product/type/H12700B/index.html} and readout electronics modules was done at JLab, including a battery of tests using a dedicated laser test setup which illuminates the MAPMTs to measure the single photoelectron response for each pixel individually to determine the gain and efficiency, following the procedure described in Ref.~\cite{Contalbrigo:2020snw}.

\begin{figure} [h]
    \includegraphics[width=0.5\textwidth]{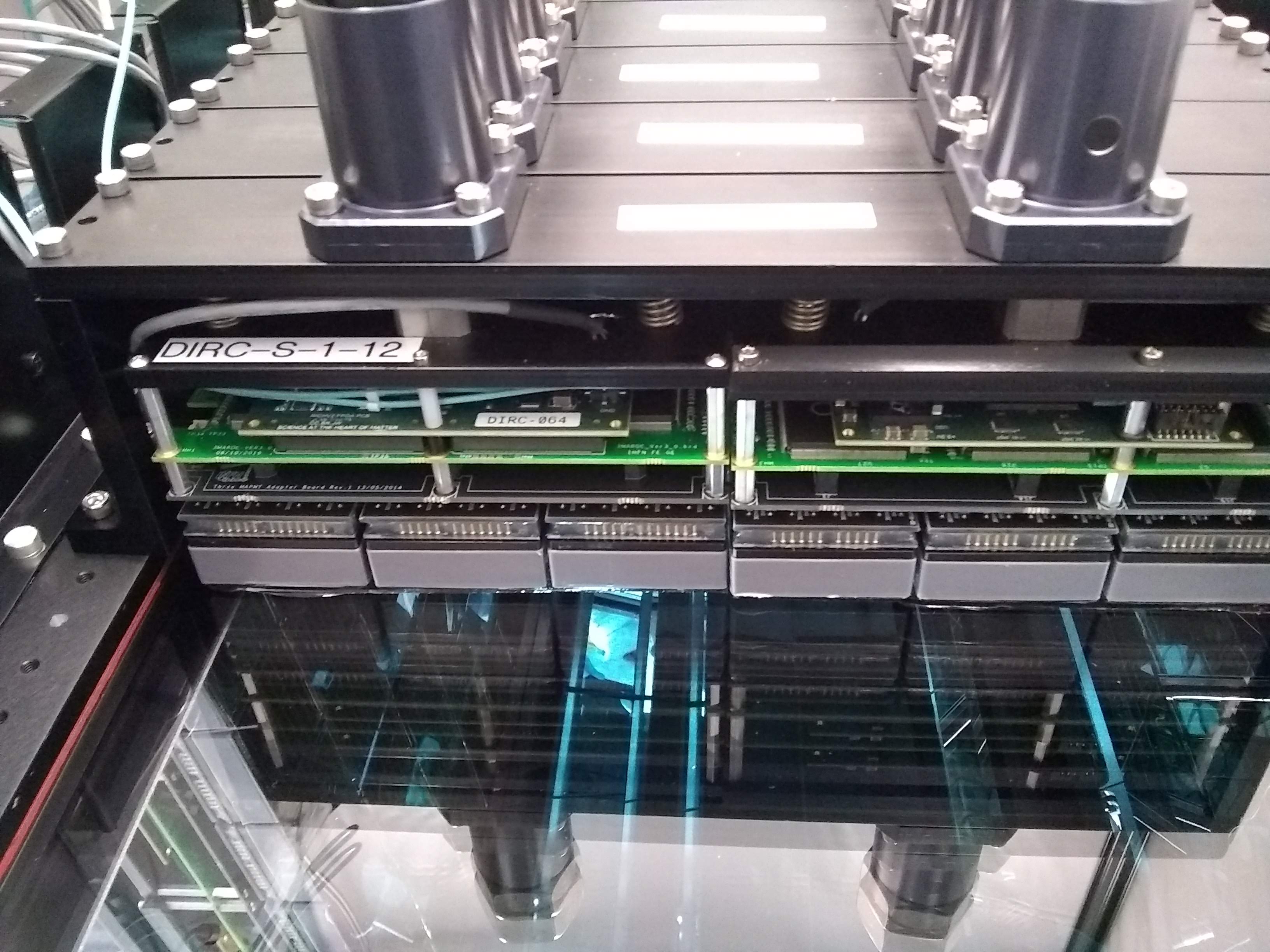}
    \includegraphics[width=0.485\textwidth]{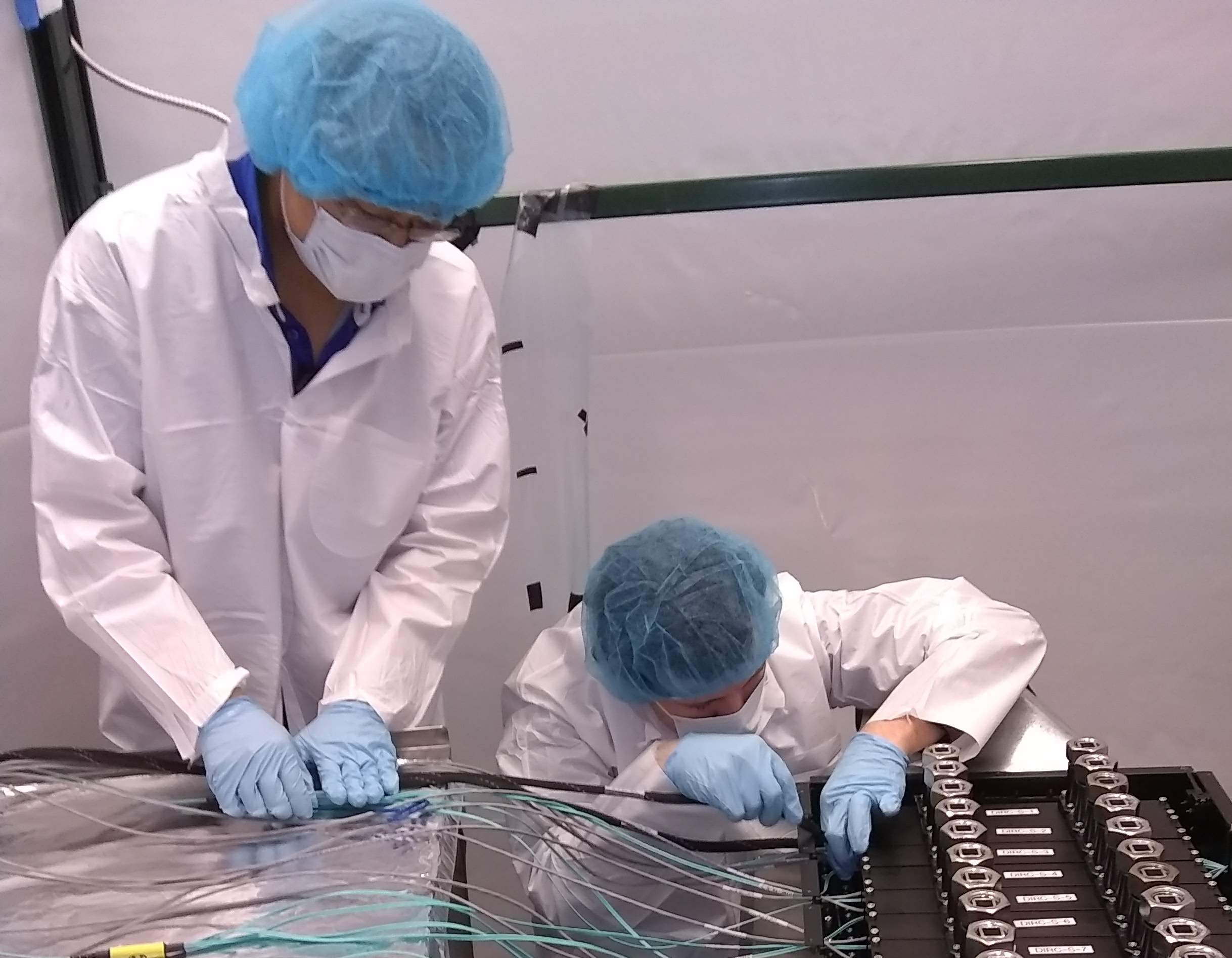}
    \caption{Installation of DIRC MAPMT modules on fused silica window of the Optical Box (left) and cabling of modules to complete the installation (right).}
    \label{fig:pmt}
\end{figure}

In preparation for the first commissioning beamtime, half of the assembled and tested MAPMT modules were then installed for one of the Optical Boxes, as shown in Fig.~\ref{fig:pmt} (left).  The MAPMTs are coupled to the fused silica window of the OB by a silicone cookie and compressed by the black bracket seen in the figure~\cite{Patsyuk:2020wnh}.  Each module has its own low and high voltage as well as optical fiber, which were installed and routed through the dark box as seen in Fig.~\ref{fig:pmt} (right).  The fully-instrumented OB was coupled to the bar boxes with a water-tight gasket, followed by a complete test of the readout electronics prior to the commissioning beamtime, using the LED calibration system.

\section{Spring 2019 Commissioning}
\label{sec:spring}

\begin{figure} [h]
    \centering
    \includegraphics[width=0.99\textwidth]{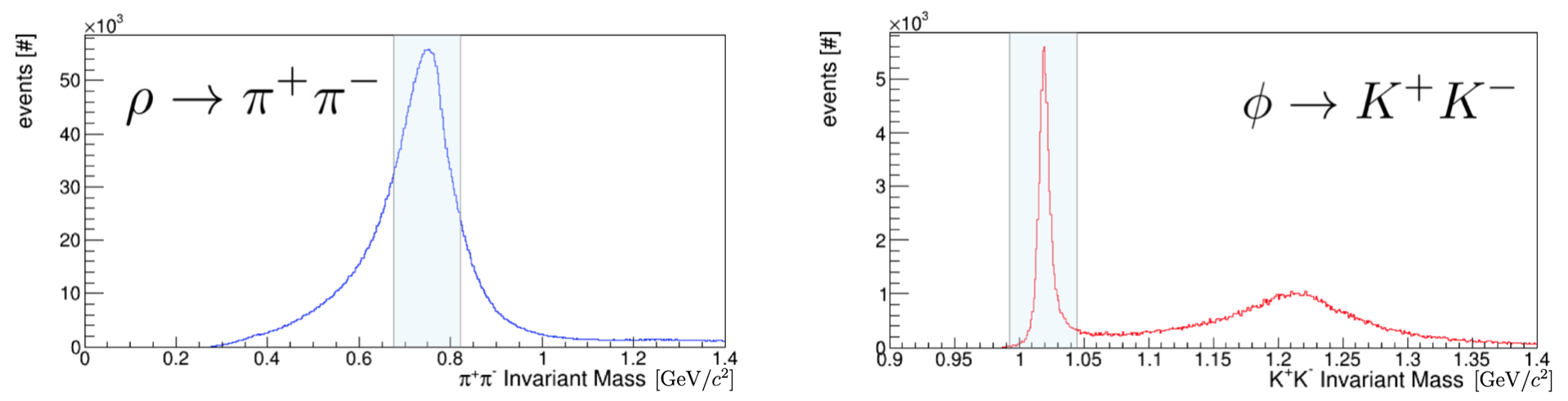}
    \caption{Invariant mass of exclusively produced $\pi^+\pi^-$ (left) and $K^+K^-$ (right) showing clear $\rho$ and $\phi$ peaks selected to study the DIRC detector performance.}
    \label{fig:rhophi}
\end{figure}

Over a ten-day period in February 2019, half of the \gx{} DIRC was commissioned by operating the nominal \gx{} experiment with the addition of one Optical Box readout with the standard DAQ system with an open trigger provided by coincidence of minimal energy deposits in the forward and barrel calorimeters.  The nominal \gx{} experiment provided the trigger and final state particle reconstruction, yielding high statistics samples of exclusive $\gamma p \rightarrow \rho p,  \rho \rightarrow \pi^+\pi^-$ and $\gamma p \rightarrow \phi p,  \phi \rightarrow K^+K^-$ events as shown by the invariant mass distributions in Fig.~\ref{fig:rhophi} (left) and (right), respectively.  These event samples provided pure samples of $\pi^\pm$ and $K^\pm$ track samples to test the DIRC reconstruction algorithms and particle identification performance.  

\begin{figure} [h]
    \centering
    \includegraphics[width=0.55\textwidth]{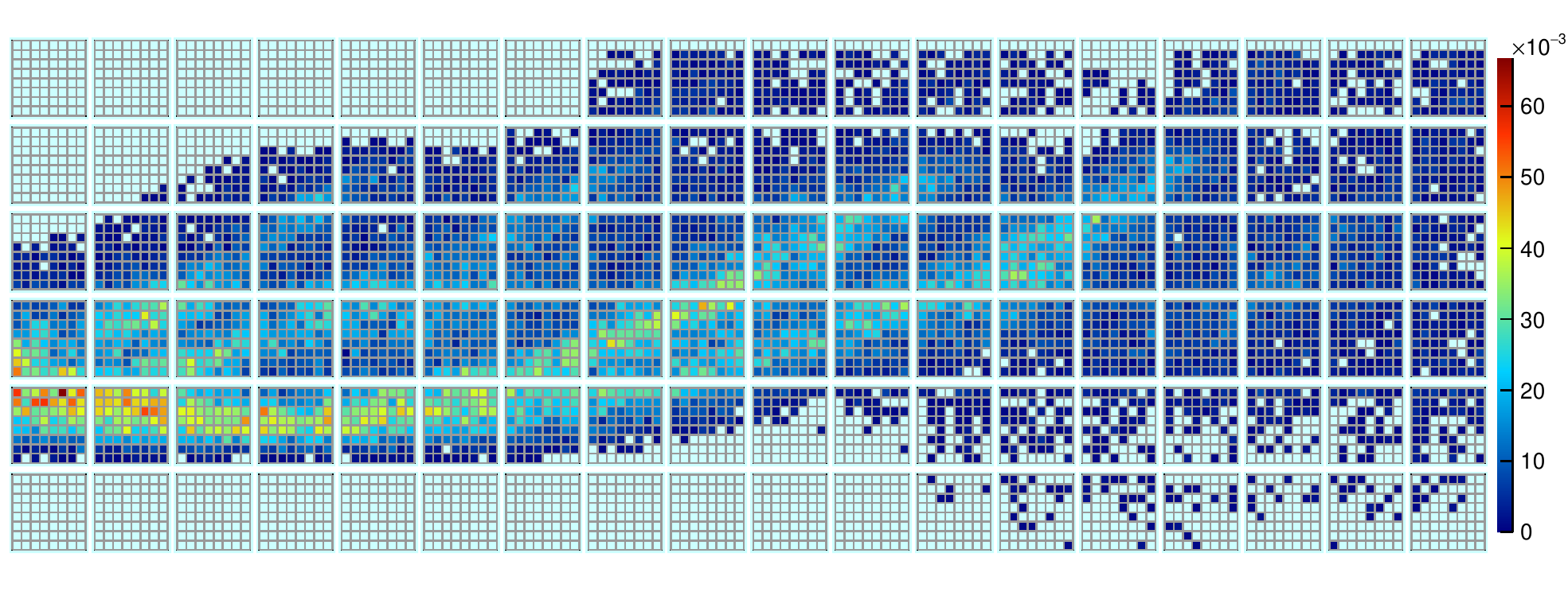}
    \includegraphics[width=0.55\textwidth]{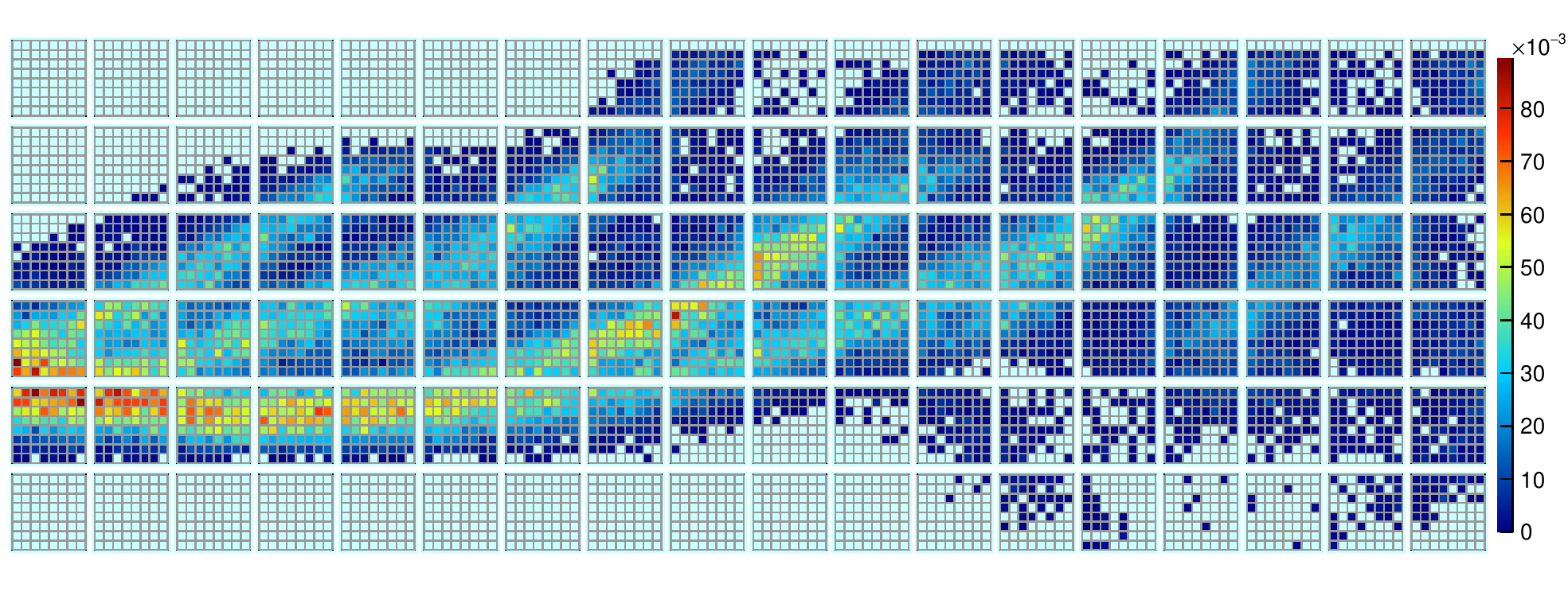}
    \caption{DIRC Cherenkov photon hit unnormalized hit intensity over MAPMT plane for identified $\pi^+$ tracks, comparing data (top) with expected distribution from GEANT MC simulation (bottom).}
    \label{fig:hitpattern}
\end{figure}

The occupancy of the MAPMT pixels form a 2-dimensional array  (48 columns by 144 rows) where the Cherenkov photons from the DIRC radiators are imaged as shown in Fig.~\ref{fig:hitpattern} for a sample of selected $\pi^+$ tracks.  The tracks from the data (described above) are shown on the top panel of Fig.~\ref{fig:hitpattern} while the bottom panel shows the analogous distribution from a GEANT simulation sample of selected $\pi^+$ tracks.  These occupancy distributions were obtained in nearly real-time during the commissioning, and their consistency was critical to confirm the validity of the data being recorded as well as the implementation of the simulated detector geometry.

\begin{figure} [h]
    \centering
    \includegraphics[width=0.48\textwidth]{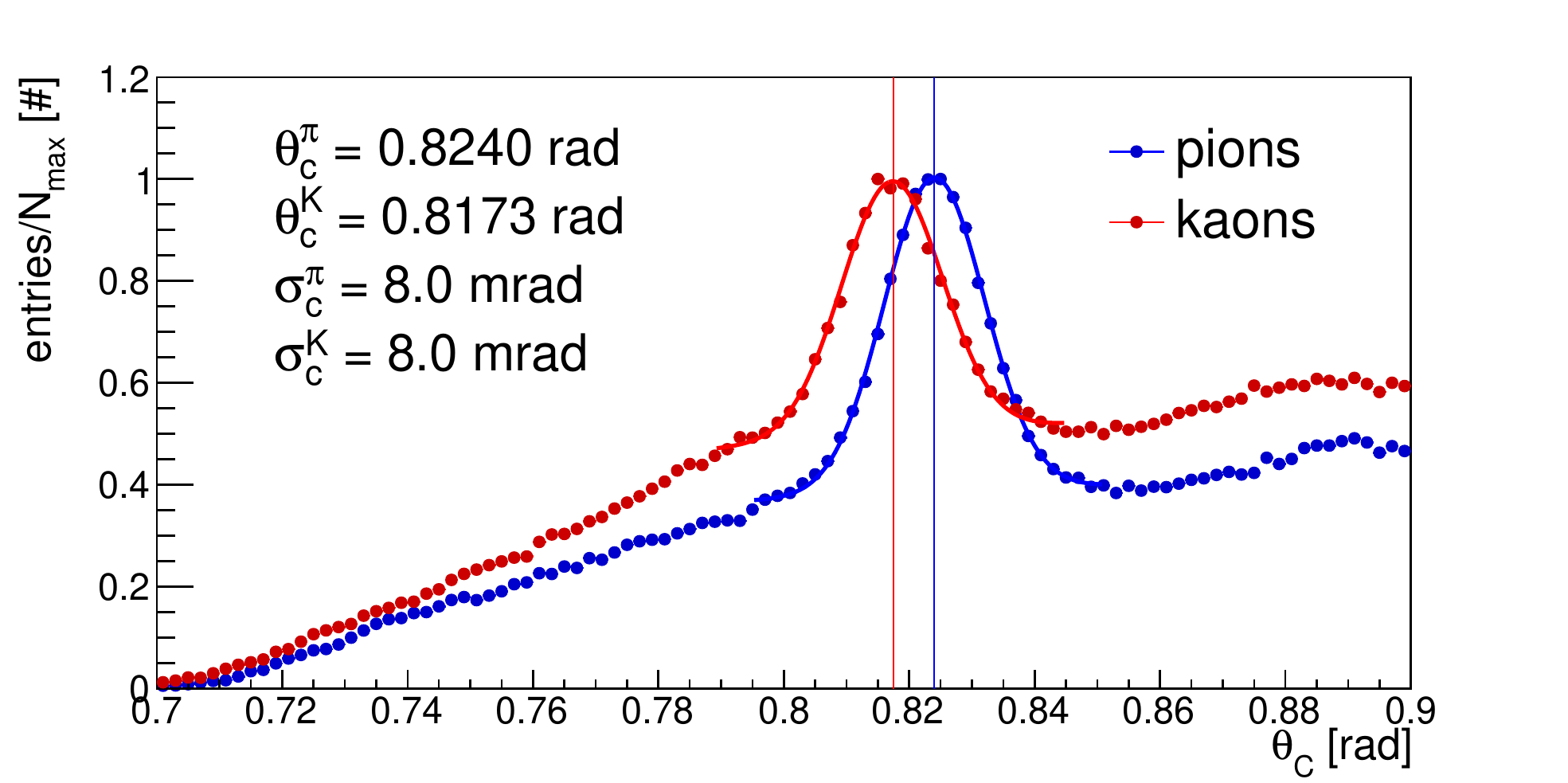}
    \includegraphics[width=0.48\textwidth]{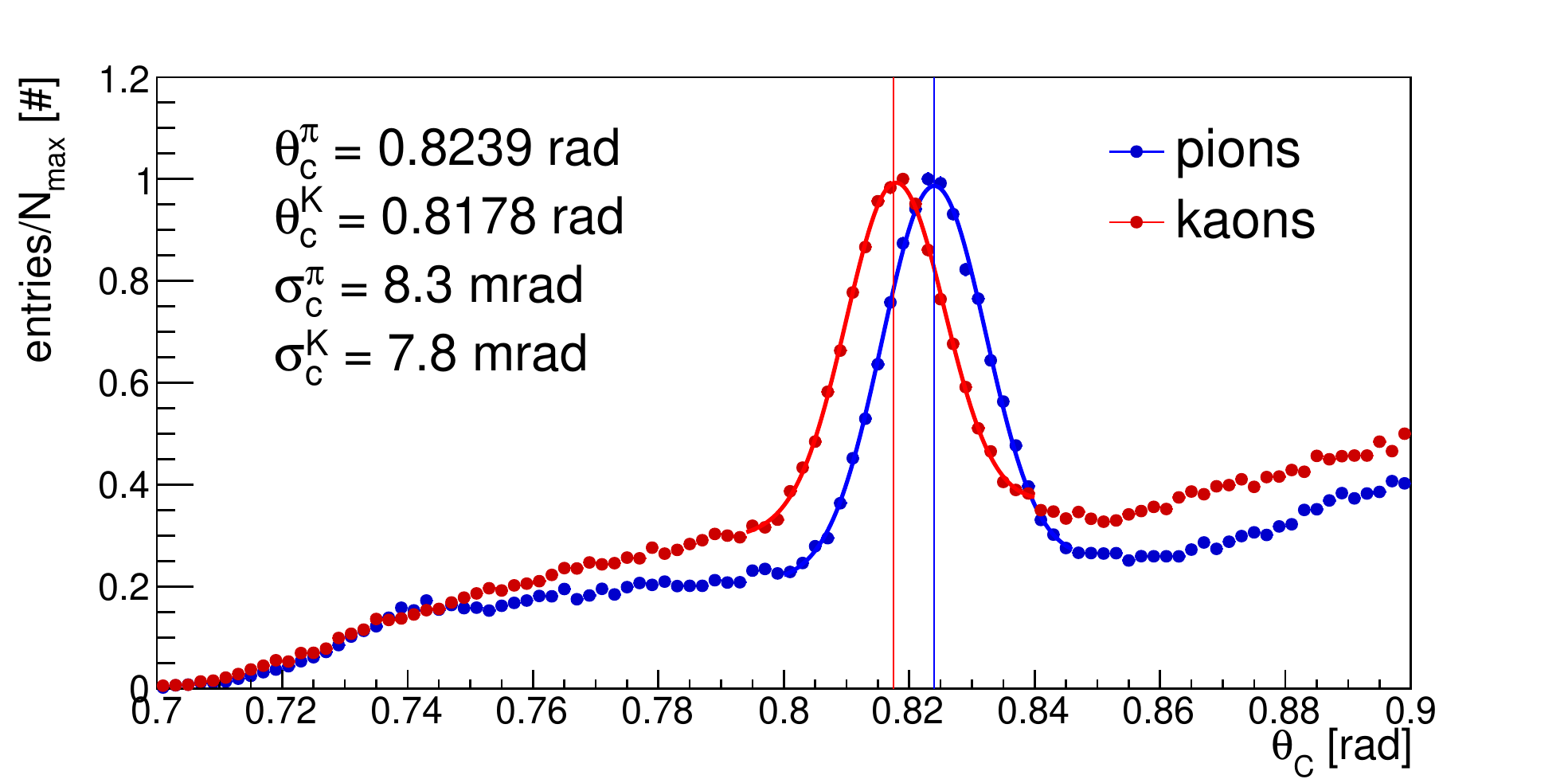}
    \caption{Cherenkov angle $\theta_C$ distribution for identified pions (blue) and kaons (red) with $p > 3.8$~GeV/$c$ identified through the $\rho$ and $\phi$ reactions described above for beam data (left) and simulation (right).}
    \label{fig:cangle}
\end{figure}

Particle identification with the DIRC is performed by comparing the observed Cherenkov photon pattern for a single track to the expected Cherenkov angles for a $\pi$ and $K$ mass hypothesis, and computing the likelihood for each hypothesis assuming a Gaussian resolution on the Cherenkov angle.  The measured Cherenkov angles per photon for charged pions and kaons are shown in Fig.~\ref{fig:cangle} for $p > 3.8$~GeV/$c$ identified through the $\rho$ and $\phi$ reactions described above for beam data (left) and simulation (right).  The results shown here are from a limited range of the available polar angles, which show a good agreement between the observed mean Cherenkov angle $\theta_C$ and single photon resolution ($\sigma_C$) for the data and simulation.  The average number of detected photons per track ranges from 15 to 35, depending on the track incident angle, which is consistent with the observed photon yield at BaBar.  These results utilize preliminary calibration and alignment, which are currently under study and improvements are expected before the final performance of the DIRC can be quantified over the full available phase space.
    
\section{Fall 2019 Commissioning and Outlook}
\label{sec:fall}

After the Spring 2019 commissioning run the water was drained from the optical box to inspect the optical components.  The mirrors which had been immersed in water for $\sim$6 months were found to have degraded in some areas, where a reaction with the water reduced the reflectivity.  Therefore, in preparation for the Fall 2019 commissioning period, the mirrors were replaced in 2019 with new front surface mirrors, which were covered by a 3~mm layer of protective Borosilicate glass.  The coupling of the glass-mirror interface maintains a thin air gap between the two surfaces.

In December 2019 the complete \gx{} DIRC detector was assembled with two Optical Boxes coupled to the four bar boxes installed in Hall D.  The mirror improvements described above were implemented in both Optical Boxes and the full detector was commissioned with beam over a two week period.  Initial results show similar performance to the previous Spring 2019 commissioning period, and analysis of this data is currently underway.

\acknowledgments

We would like to acknowledge the outstanding efforts of the staff of the Accelerator and the Physics Divisions at Jefferson Lab that made the experiment possible.  This work is supported by the U.S. Department of Energy, Office of Science, Office of Nuclear Physics under contracts DE-AC05-06OR23177, DE-FG02-05ER41374 and Early Career Award contract DE-SC0018224 and the German Research Foundation, GSI Helmholtzzentrum f\"ur Schwerionenforschung GmbH.

\end{document}